\documentclass[twoside,slac_one]{revtex4}
\usepackage{graphicx}
\usepackage{fancyhdr}
\usepackage{amsmath} 
\usepackage{bm}
\usepackage{amsxtra}
\usepackage{amssymb}
\usepackage{amsthm}
\usepackage{latexsym}
\usepackage{lscape}

\begin{document}

\title{A Search for the Higgs Boson in the $H\rightarrow\gamma\gamma$ Channel with CMS DPF-2011 Proceedings}

%

\author{Christopher Palmer}
\affiliation{Department of Physics, University of California, San Diego, CA, USA}

\begin{abstract}
We report on a search for SM Higgs Boson in the Higgs to two photons decay channel conducted by the CMS experiment with the data accumulated during the 2010 and 2011 running of the LHC at sqrt(s) = 7 TeV.
\end{abstract}

\maketitle

\thispagestyle{fancy}


\section{Introduction - $H\rightarrow\gamma\gamma$ in a Nutshell}
We are motivated to explore this signal because of its narrow resonance over a smoothly falling background.  Despite the enormous background and the somewhat low number of expected events, this decay channel is extremely relevant to the favored, low-mass Higgs search as long as the resolution of the peak can be measured with reasonable resolution.

\begin{figure}[ht]
\centering
        \includegraphics[width=11cm]{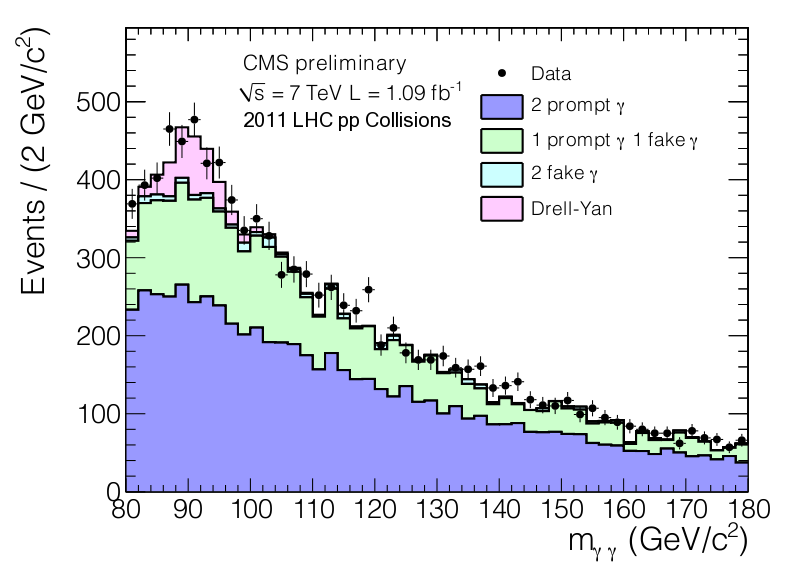} 
\caption{Final invariant mass selection with simulation.} \label{mgg}
\end{figure}

Our background can be categorized in two components:  irreducible and reducible.  Irreducible background is more difficult to eliminate because these are events which have two real photons in them.  Reducible background events are typically poorly reconstructions electrons or jets which are mistakenly taken to be photons.  In the final selection invariant mass plot (~Figure~\ref{mgg}) one can see the various components from each.

\section{Necessary CMS Design/ECAL}
\subsection{Compact Muon Solenoid (CMS)}
\begin{figure}[ht]
\centering
    \includegraphics[width=11cm]{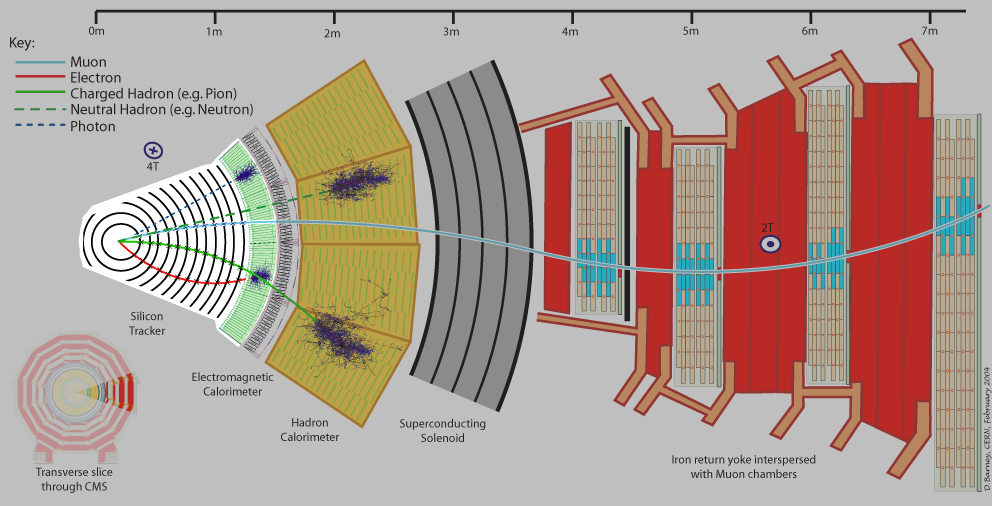}
\caption{Cross sectional schematic of CMS.} \label{cms_xsec}
\end{figure}

CMS is a general purpose detector.  It is designed to detect and reconstruct numerous physics objects (e.g. photons, electrons, muons, hadronic jets, etc).  From this general design we are compatible of seeking out new physics (e.g. Higgs and SUSY).  Below (~Figure~\ref{cms_xsec}) is a cross sectional schematic of the detectored with the relevant components highlighted (i.e. the trackers and the electromagnetic calorimeter (ECAL)).

\subsection{CMS - Electromagnetic Calorimeter (ECAL)}
The ECAL is made of $\sim$76K $PbWO_{4}$ crystals in the barrel ($|\eta|<1.48$) and the endcap ($1.48<|\eta|<3.$) of CMS.  It was reconstructed in order to determine the energy of photons and electrons to high precision.  The design resolution of unconverted photons in the barrel with energy greater than 100 GeV was $\sim$0.5\%

\begin{figure}[ht]
\centering
      \includegraphics[width=5.5cm]{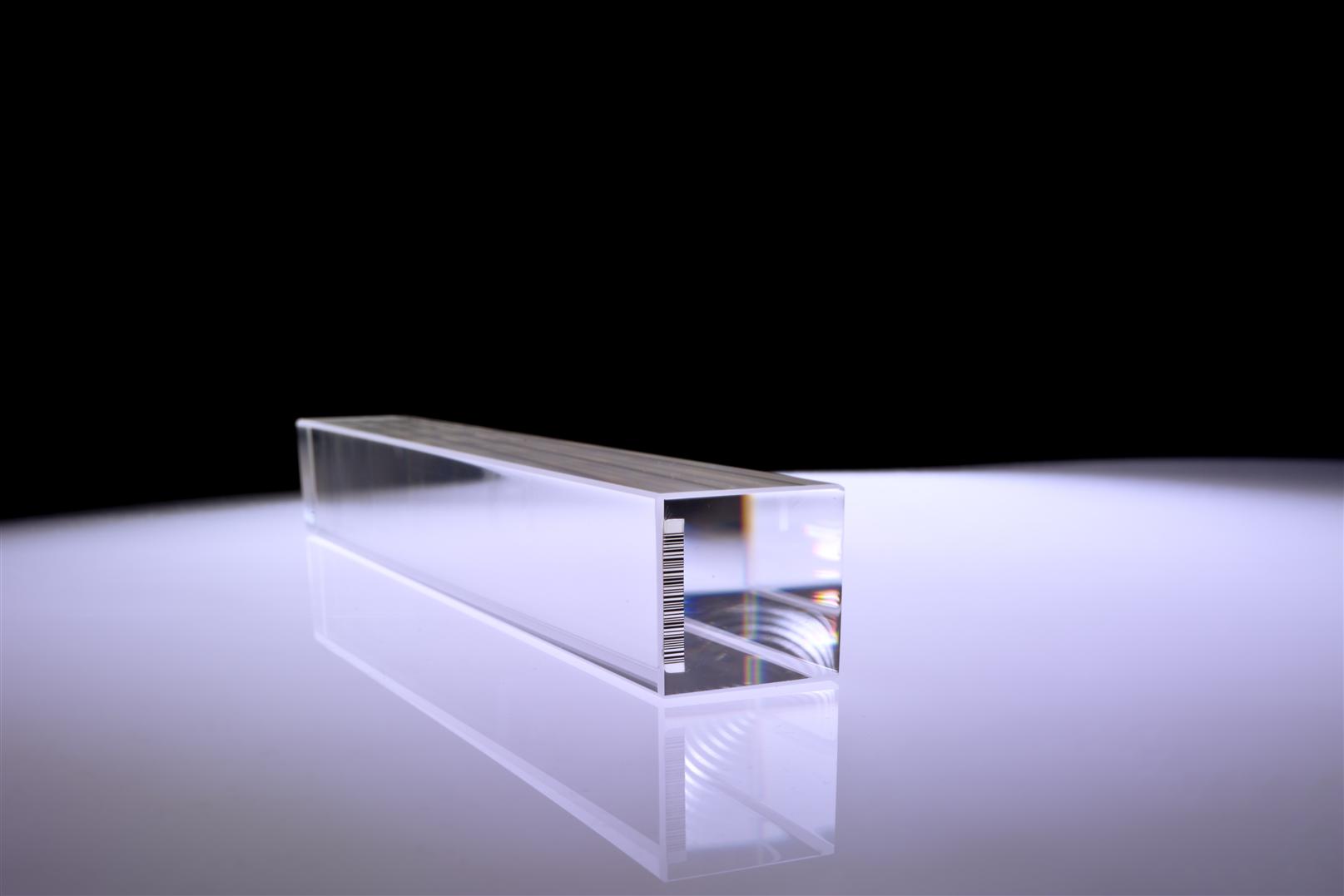} 
\caption{A photograph of one of the crystals from the ECAL.} \label{ecal_cry}
\end{figure}

\begin{figure}[ht]
\centering
      \includegraphics[width=6cm]{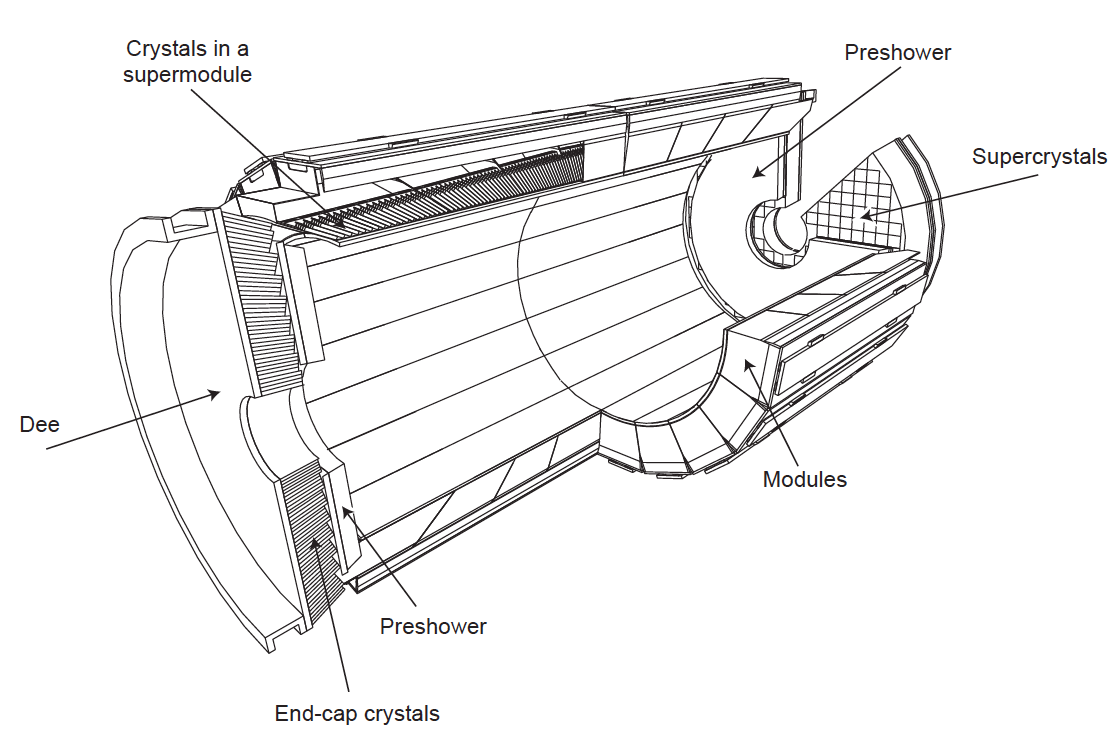}
\caption{A schmatic of CMS's ECAL.} \label{ecal_sch}
\end{figure}

There are two main critical issues that impact the resolution that CMS is able to determine. One is the calibration of the crystals.  Two techniques, which are exploited, are $\pi_{0}\rightarrow\gamma\gamma$ events for inter-crystal calibration and $Z\rightarrow e^{+}e^{-}$ for global energy scale calibration.  The other critical issue is transparency corrections for radiation damage.  An integrated laser system measures the transparency of crystals and that information is used to correct the transparency loss.  Below is a schematic of the ECAL (~Figure~\ref{ecal_sch}) and a photo of one of the crystals (~Figure~\ref{ecal_cry}).


\section{Analysis Strategy}
\subsection{DiPhoton Selection}
\begin{figure}[ht]
\centering
        \includegraphics[width=8cm]{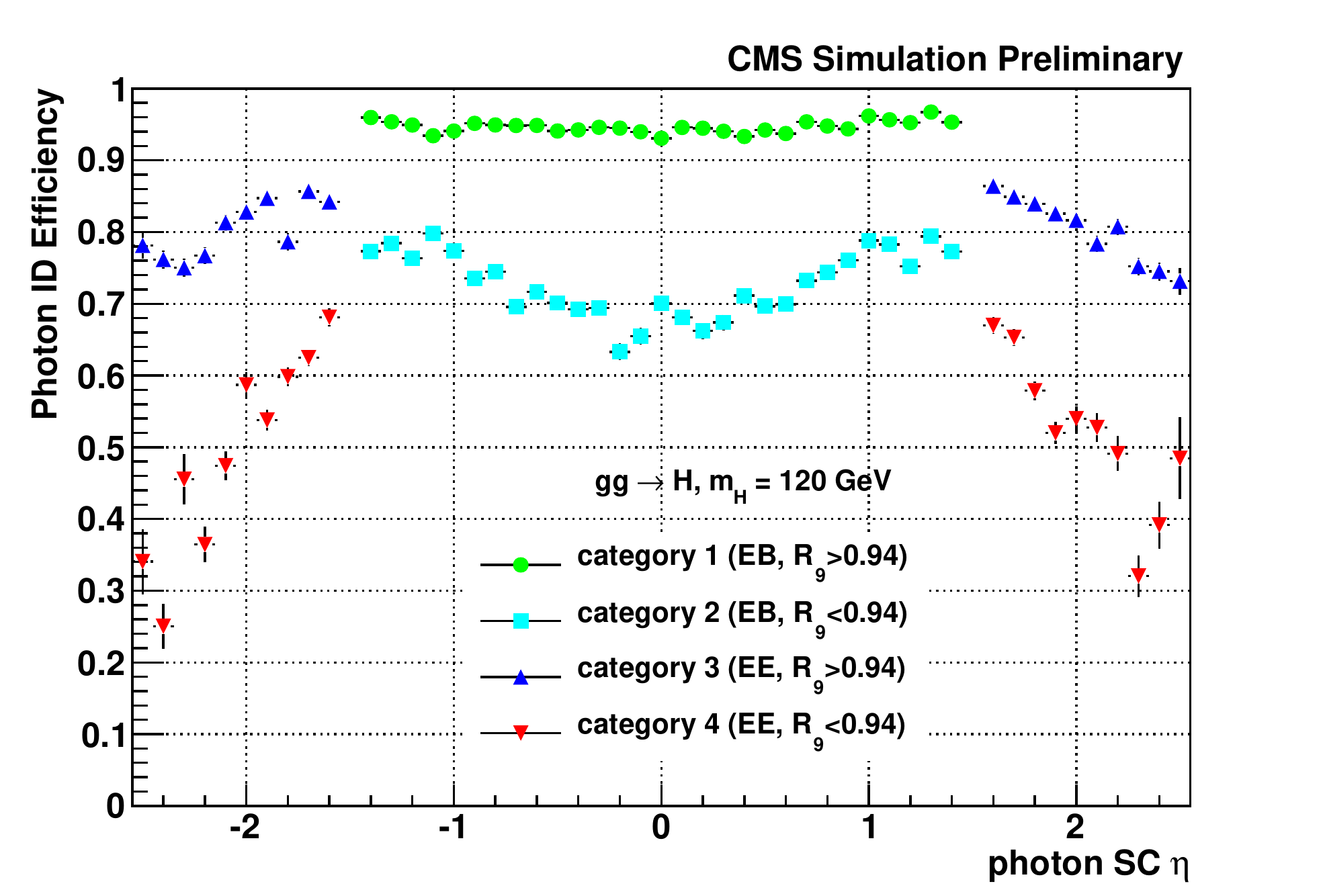} 
\caption{Efficiency of selection in categories as a function of $\eta$ with simulated Higgs' photons.} \label{mc_eff}
\end{figure}

Photon identification is important for removal of reducible background.  The primary tools used to distinguish between signal and background are isolation, cluster shape and electron veto.  The cuts set in these photon identification variables are set separately for in categories defined by the photon being in the barrel or one of the endcaps and by whether or not the photon converted to an electron-positron before before reaching the ECAL.  The efficiency versus $\eta$ is plotted below for these four categories (~Figure~\ref{mc_eff}).  Also, note that our selection contains minimum thresholds on the transverse momentum of the two photons (40 and 30 GeV).

\subsection{Vertex Selection}
In our data, numerous proton-proton interactions occur each time bunches of protons pass through each other.  If the wrong vertex is chosen to reconstruct the photons then the energy of the photons and the mass of the di-photon will be incorrect.  Effectively smearing the mass of the Higgs signal.
        
\begin{figure}[ht]
\centering
        \includegraphics[width=4.3cm]{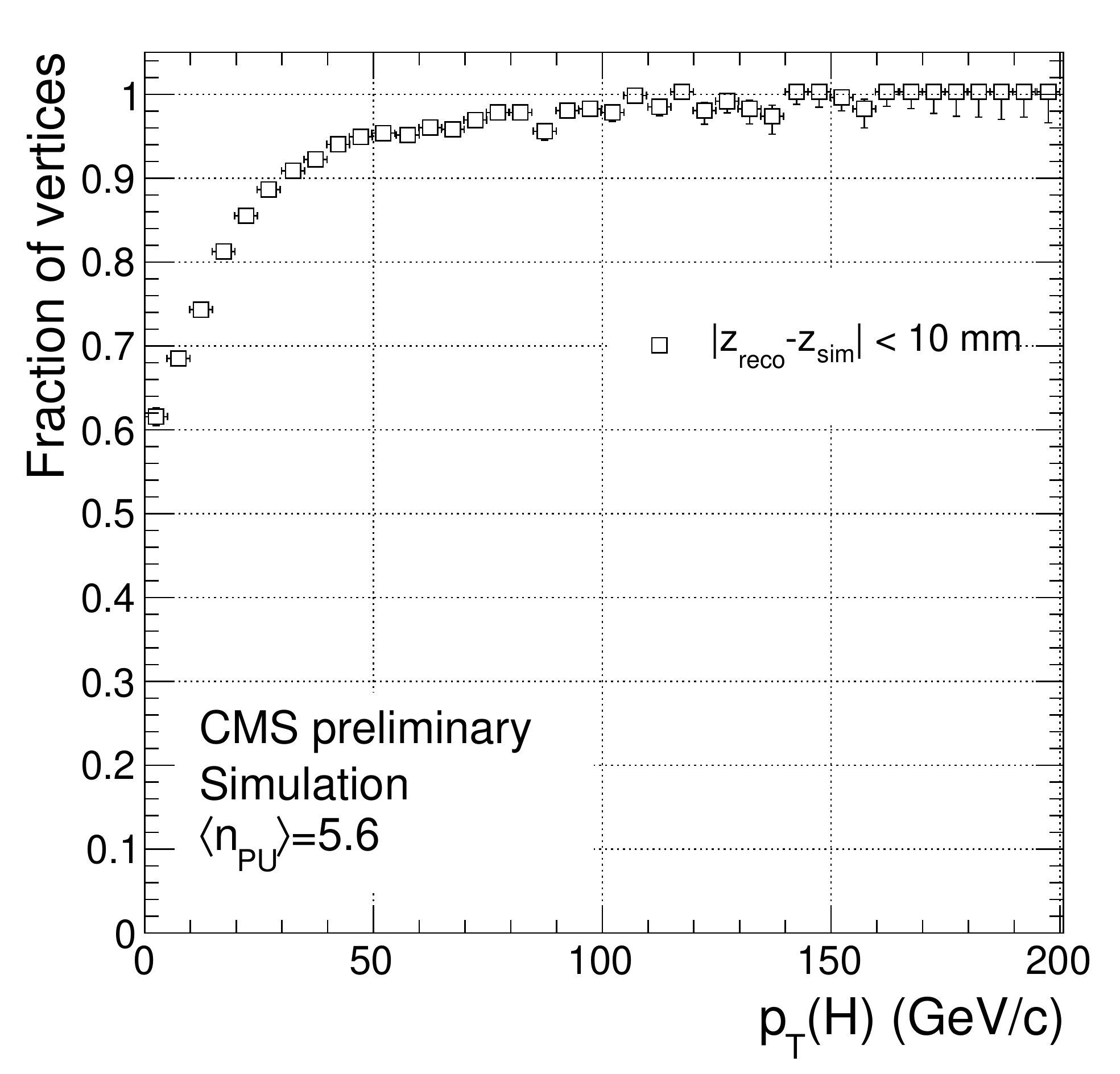}
\caption{Efficiency of vertex selection within 1 cm as function of Higgs' transverse momentum.} \label{vertex_eff}
\end{figure}

We use the following quantities to increase our probability of determining the correct vertex. For unconverted photons we use $\displaystyle\sum\limits_{Tracks} P_{T}^{2}$, projection of tracks onto $\gamma\gamma$ and the balance between $\gamma\gamma$ and vertex's tracks.  For converted $\gamma$'s the conversion-tracks are used to point back to vertex.  The efficiency of this method is plotted below as a function of di-photon transverse momentum (~Figure~\ref{vertex_eff}).

\section{Efficiencies from Data}
Making the basic assumption that electrons and photons have similar shower properties we use the tag and probe method with $Z\rightarrow e^{+}e^{-}$ electrons to determine the efficiency of our photon selection.  First events with robustly identified electrons are selected and we seek an additional electron passing a minimal transverse energy requirement, the probe.  The photon reconstructed with the same energy deposit has the photon cuts applied to it and efficiency of passing events to selected events is measured in each category.  Below is a table of these efficiencies.  Please note the majority of our photons are in the unconverted, barrel category, which is more than 90\% efficient.

To measure the efficiency of the electron veto cuts, photons from $Z\rightarrow \mu^{+}\mu^{-}\gamma$ events have the selection applied.  Efficiencies are listed in the table below.

\begin{table}[ht]
\begin{center}
\caption{Tag and Probe Efficiency}
    \begin{tabular}{| c | c | c | c|}
          \hline
          Category &
          $\epsilon_{data}$ (\%) & $\epsilon_{MC}$ (\%) & $\epsilon_{data}$/$\epsilon_{MC}$ \\
          \hline
          \multicolumn{4}{|c|}{\tiny All cuts except electron rejection (from $Z\rightarrow e^{+}e^{-}$)}\\
          \hline
          1 & 91.77$\pm$0.14 & 92.43$\pm$0.07 & 0.993$\pm$0.002 \\
          2 & 72.67$\pm$0.43 & 71.89$\pm$0.08 & 1.011$\pm$0.007 \\
          3 & 80.33$\pm$0.47 & 80.04$\pm$0.18 & 1.004$\pm$0.008 \\
          4 & 57.80$\pm$1.26 & 55.09$\pm$0.15 & 1.049$\pm$0.025 \\
          \hline
          \multicolumn{4}{|c|}{\tiny Electron rejection cut (from $Z\to\mu\mu\gamma$)}\\
          \hline
          1 & 99.78$^{+0.13}_{-0.16}$ & 99.59$^{+0.13}_{-0.17}$ & 1.002$^{+0.002}_{-0.002}$ \\
          2 & 98.77$^{+0.59}_{-0.73}$ & 97.70$^{+0.32}_{-0.37}$ & 1.011$^{+0.007}_{-0.008}$ \\
          3 & 99.32$^{+0.51}_{-1.02}$ & 99.29$^{+0.30}_{-0.42}$ & 1.000$^{+0.006}_{-0.011}$ \\
          4 & 93.0$^{+2.1}_{-2.3}$ & 93.34$^{+0.79}_{-0.86}$ & 0.996$^{+0.024}_{-0.027}$ \\
          \hline
        \end{tabular}
\label{example_table}
\end{center}
\end{table}

\begin{table}[ht]
\begin{center}
\caption{Electron Veto Efficiency}
        \begin{tabular}{| c | c |}
          \hline
          \multicolumn{2}{|c|}{Both photons in barrel}\\  
          \hline
          2 Unconverted & 1,2 Converted\\
          \hline
          100.00$^{+0.00}_{-0.01}$\% & 99.53$\pm$0.04\%\\ 
          \hline
          \hline
          \multicolumn{2}{|c|}{One or more in endcap}\\
          \hline
          2 Unconverted & 1,2 Converted\\
          \hline
          100.00$^{+0.00}_{-0.02}$\% & 98.86$\pm$0.07\%\\
          \hline
        \end{tabular}
\label{example_table}
\end{center}
\end{table}

\section{Resolution Results}
\begin{figure}[ht]
\centering
    \includegraphics[width=5.5cm]{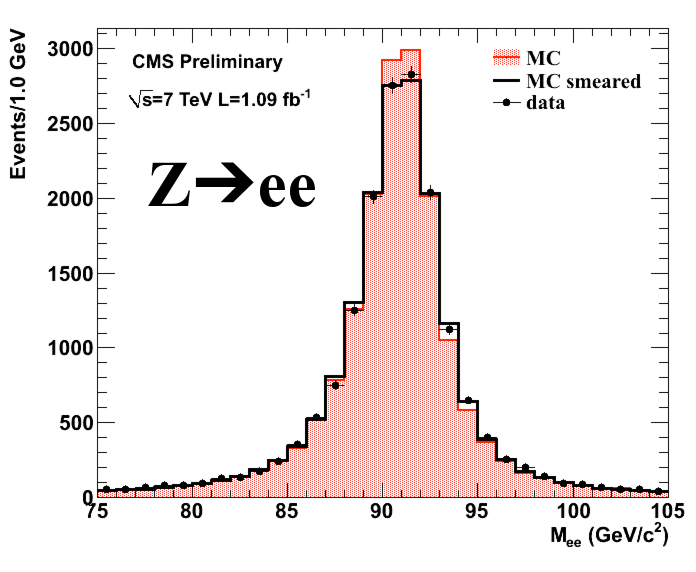} 
    \caption{Resolution from $Z\rightarrow e^{+}e^{-}$ events in data.} \label{Zee_res}
\end{figure}

\begin{figure}[ht]
\centering
        \includegraphics[width=5.5cm]{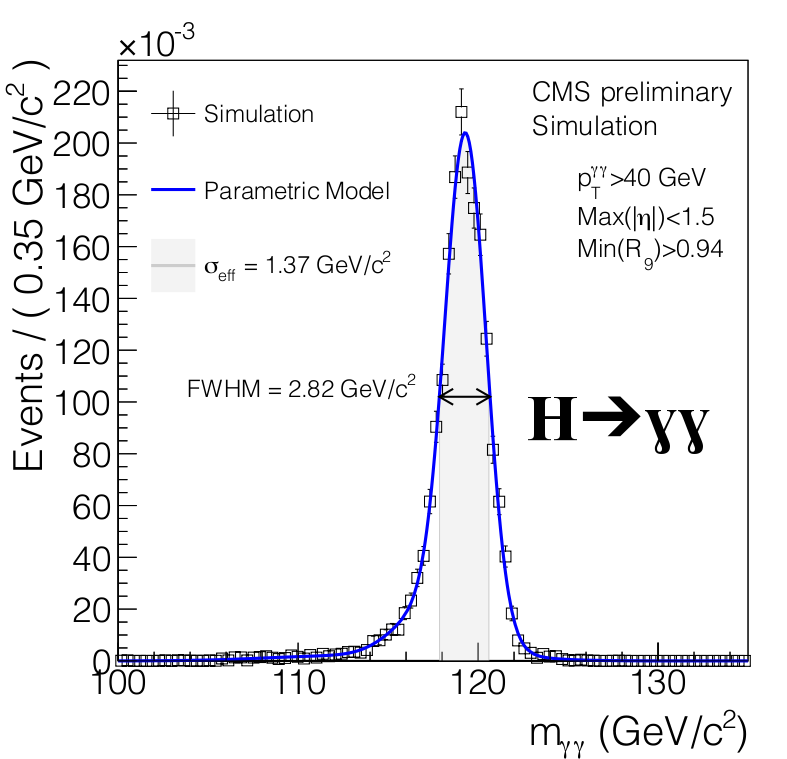}     
    \caption{Signal resolution in $H\rightarrow\gamma\gamma$ estimated from data.} \label{Hgg_res}
\end{figure}

ECAL resolution measured from $Z\rightarrow e^{+}e^{-}$ (~Figure~\ref{Zee_res})is applied to simulated Higgs' $\gamma$'s (~Figure~\ref{Hgg_res}).  The simulated Higgs' $\gamma$'s with data resolution are used in signal modeling for CL limits.  Suboptimal transparency loss corrections may be responsible for degraded resolution. An example of the measured resolution and its application to the best Higgs' photons category are below.

\section{Cateogories and Limit Setting}
\subsection{Event Classes Used for CL Evaluation}
There are eight event classes in which confidence levels are computed.  There are both photons in the barrel or either in an endcap times both converted or either unconverted times high/low diphoton transverse momentum classifications.  Below is a table which contains the fraction of signal and background in each event class.

\begin{table}[ht]
\begin{center}
    \begin{tabular}{|l|c|c|c|c|}
        \hline
        & \multicolumn{2}{c|}{Both $\gamma$'s in barrel} & \multicolumn{2}{c|}{One or more in endcap}\\
        & 2 Unconverted & 1,2 Converted & 2 Unconverted & 1,2 Converted \\
        \hline
        {\bf $\mathbf{P_{T}^{\gamma\gamma} <}$ 40 GeV/c} & & & & \\
        Signal & 0.209 & 0.271 & 0.094 & 0.116\\
        Background & 0.167 & 0.263 & 0.129 & 0.203\\
        \tiny Signal $\sigma_{effective}$ (GeV/$c^{2}$) & 1.58 & 2.33 & 3.14 & 3.60 \\
        \hline
        {\bf $\mathbf{P_{T}^{\gamma\gamma} >}$ 40 GeV/c} & & & & \\
        Signal & 0.102 & 0.122 & 0.035 & 0.051\\
        Background & 0.043 & 0.079 & 0.043 & 0.074\\
        \tiny Signal $\sigma_{effective}$ (GeV/$c^{2}$) & 1.37 & 2.12 & 2.95 & 3.26\\
        \hline
    \end{tabular}  
\label{example_table}
\end{center}
\end{table}

\begin{figure}[ht]
\centering
        \includegraphics[width=5.5cm]{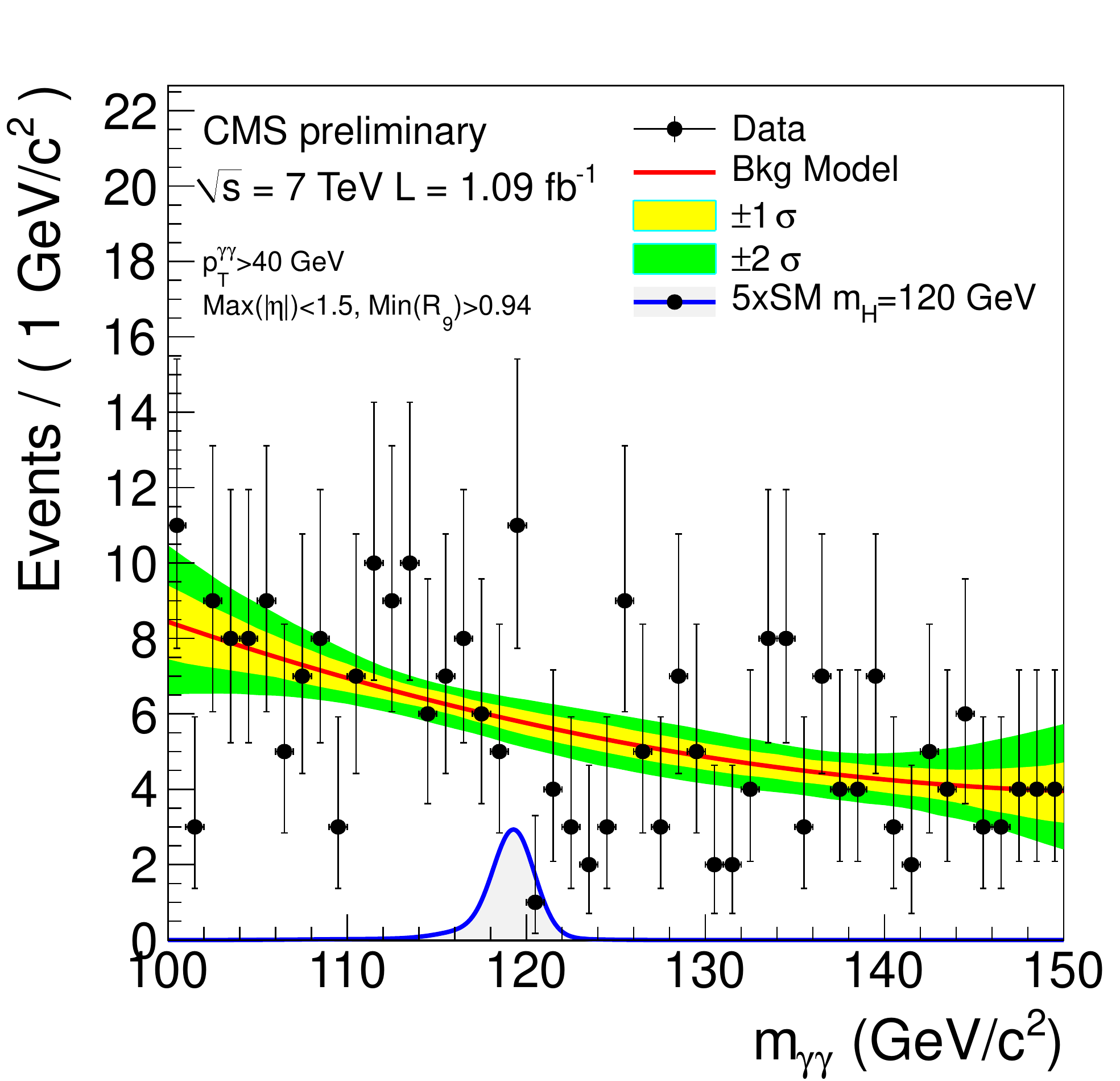} 
    \caption{High $P_{T}^{\gamma\gamma}$, both high $R_{9}$, barrel-barrel .} \label{datafit1}
\end{figure}

\begin{figure}[ht]
\centering
        \includegraphics[width=5.5cm]{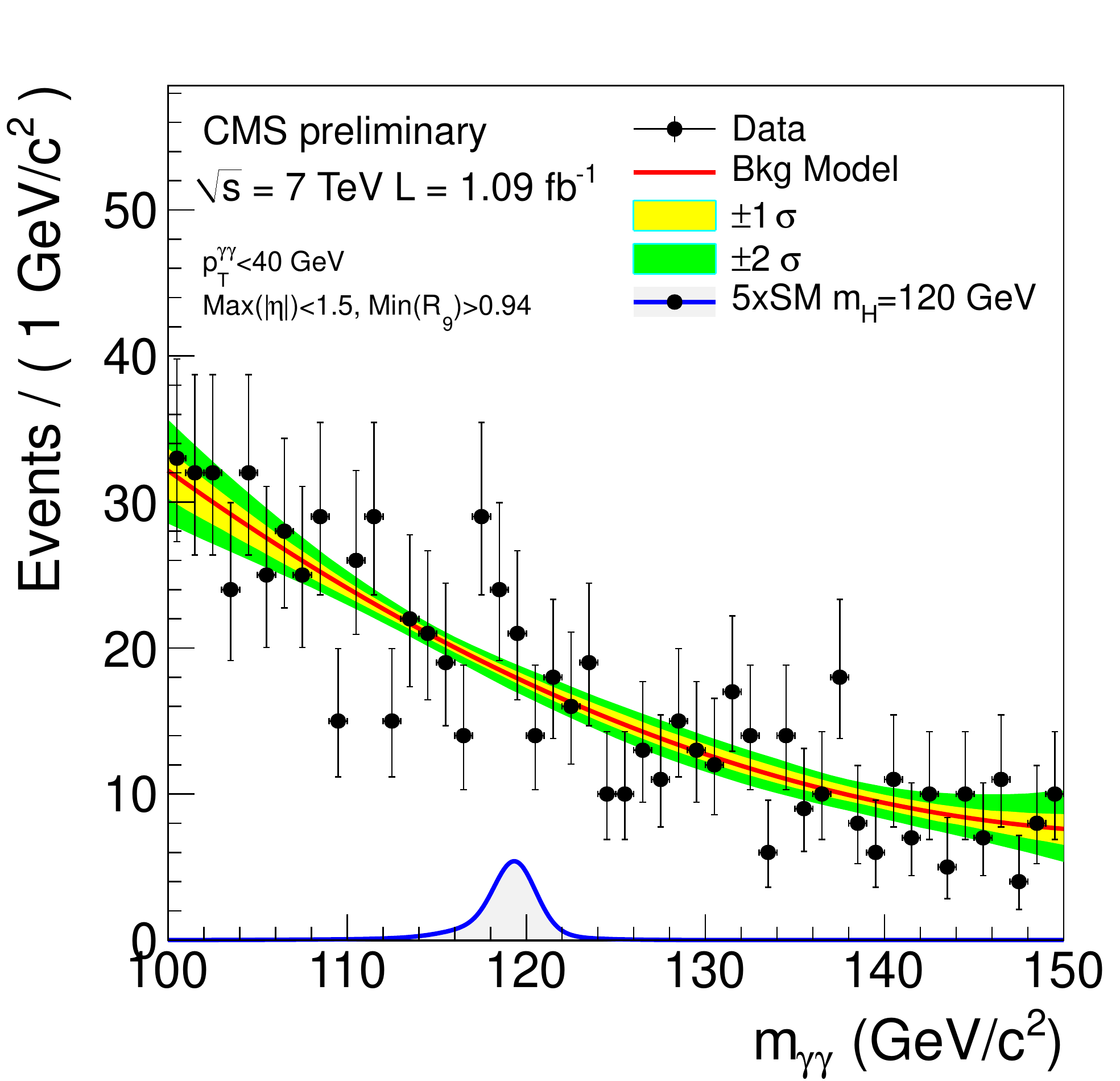} 
    \caption{Low $P_{T}^{\gamma\gamma}$, both high $R_{9}$, barrel-barrel .} \label{datafit2}
\end{figure}

Second order polynomial fits are performed in each of the eight categories on the data.  The resulting fit is the background distribution used for toy experiments.  The two both barrel, both unconverted categories are below (~Figure~\ref{datafit1} and ~Figure~\ref{datafit2}).

\subsection{Systematics}
\begin{figure}[ht]
\centering
    \includegraphics[width=11.cm]{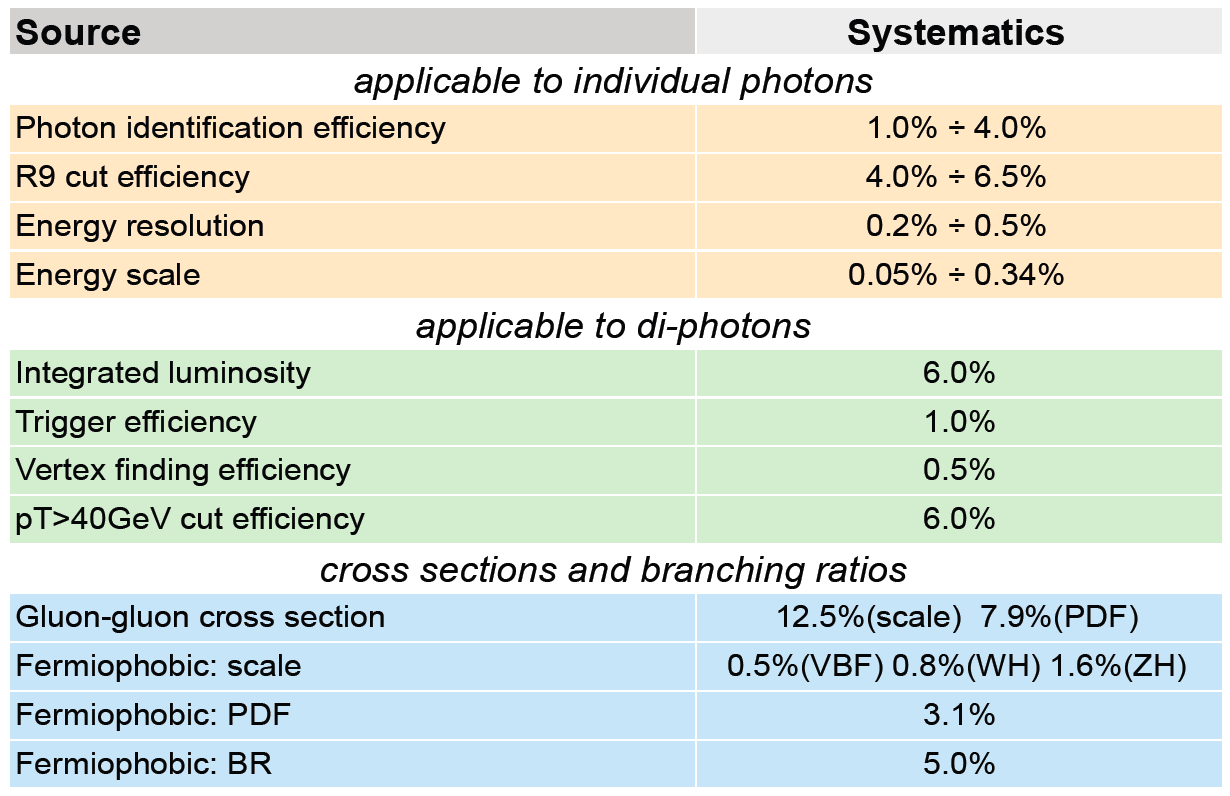} \label{syst}
\end{figure}

Below (~Figure~\ref{syst}) is a table that summarizes the systematics that are applied to the signal models.  Since the background model is from data, no systematics are applied to it.

\subsection{Evaluated Limits}
\begin{figure}[ht]
\centering
    \includegraphics[width=9cm]{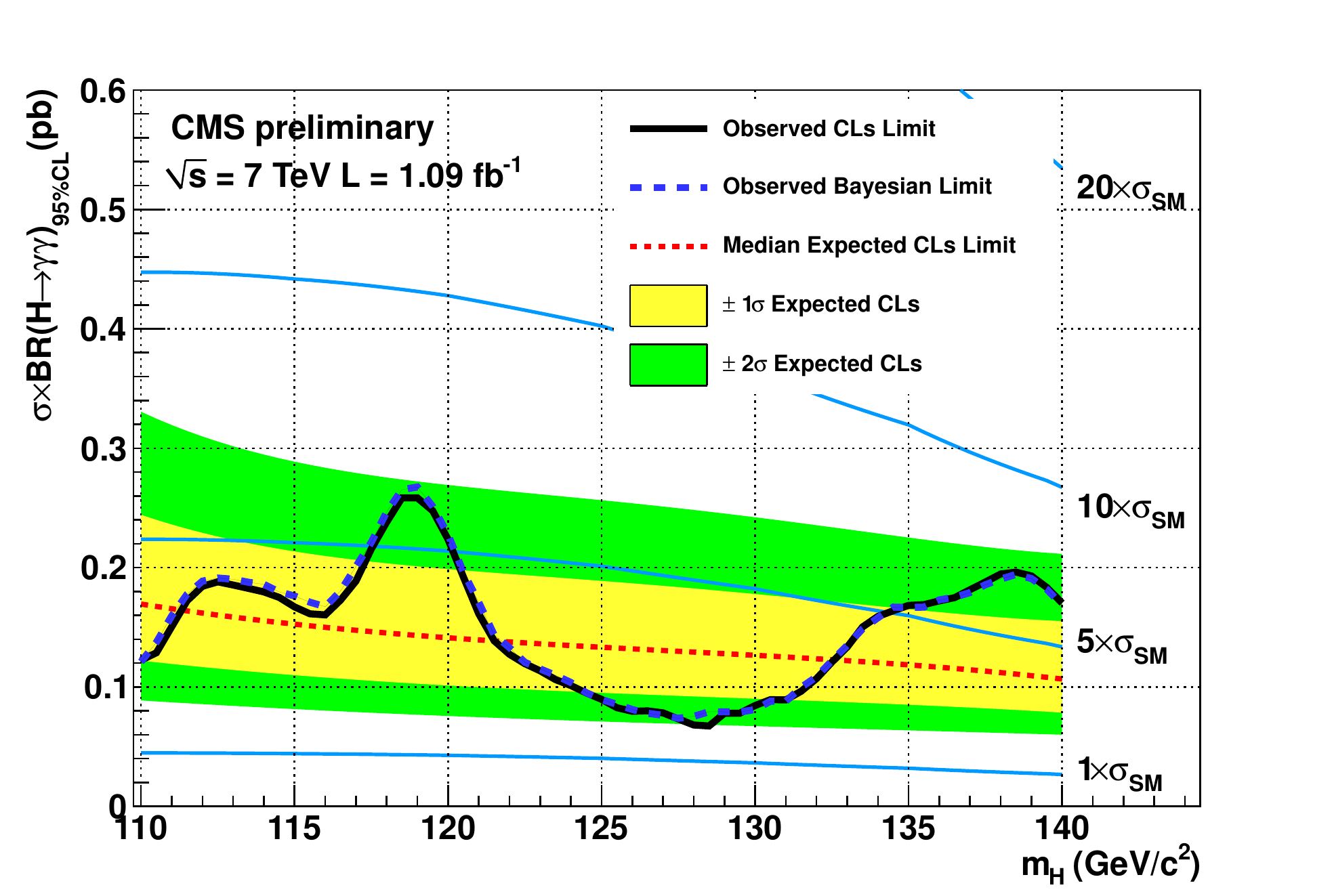} \label{sm_ex}
\end{figure}

\begin{figure}[ht]
\centering
    \includegraphics[width=9cm]{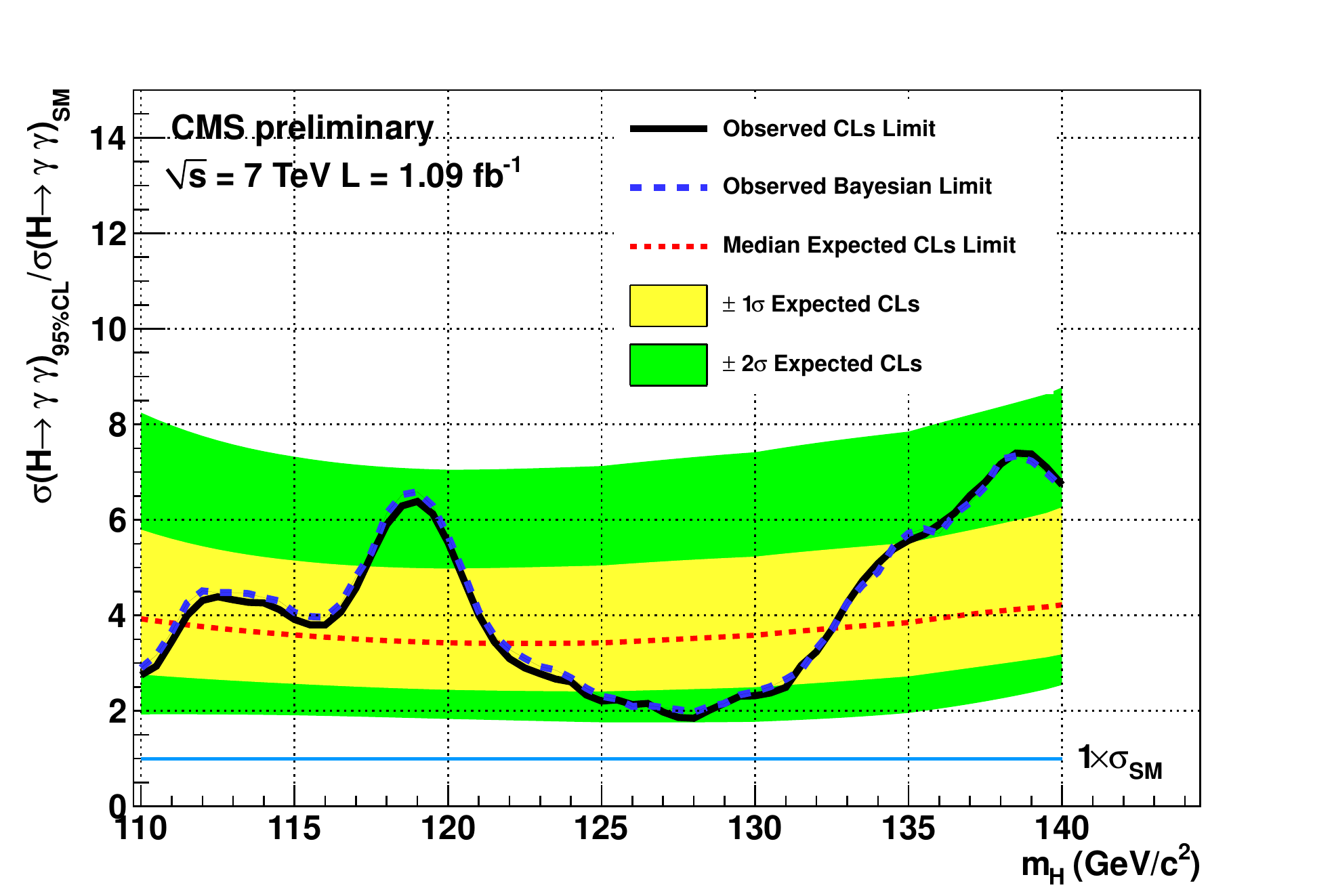} \label{sm_ex_rel}
\end{figure}

\begin{figure}[ht]
\centering
    \includegraphics[width=9cm]{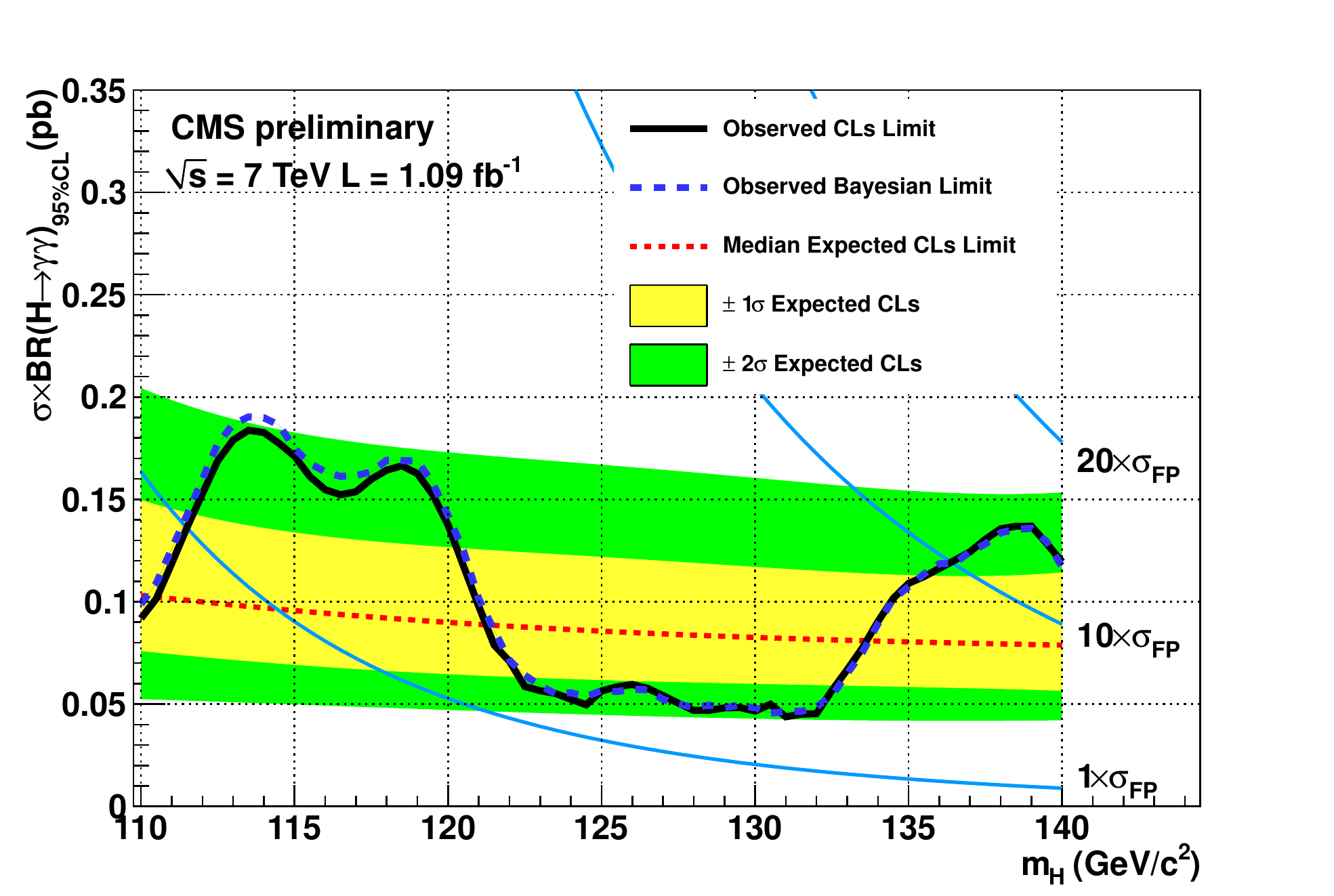} \label{fp_ex}
\end{figure}

Confidence Level (CL) limits are determined in two ways with extremely consistent results.  Our official limits are set with the modified frequentist approach (CLs) using profile likelihood.  As a cross check with all use the Bayesian method with flat prior in cross section.  In the following two plots show SM exclusions between cross sections of between 0.06 and 0.26 pb (~Figure~\ref{sm_ex}) and between 1.9 and 7.0 $*\sigma_{SM}$ (~Figure~\ref{sm_ex_rel}).  While the following plot shows Fermiophobic Higgs cross section excluded between 0.04 and 0.18 pb (~Figure~\ref{fp_ex}) and its mass constrained to be greater than 111 GeV.

\section{Conclusions/Outlook}
Our analysis is defined by our use of photon selection in categories and our use of various methods to select the best vertex possible.  Data is used to determine the resolution of our signal (via $Z\rightarrow e^{+}e^{-}$ events) model and our background model is determined directly from the fit of the data in event classes.  The CL evaluation in event classes improves our sensitivity to both the SM and Fermiophobic Higgs.

We are keen to improve this analysis by taking as much data as possible and by improving the resolution of CMS’s ECAL.  Improving resolution could be the signal greatest improvement to the SM analysis and work is ongoing in this field.

\begin{acknowledgments}
Vivek Sharma and Christoph Paus have put forth great effort in coordinating the Higgs effort within CMS over the past year.  The Higgs result from CMS would not have been achieved without them.  Christopher Seez and Paolo Meridiani have been directing the research specifically outlined in this proceeding and should be commended for their efforts and the group's achievements.  Finally, I would like to personally thank my advisor Jim Branson and Marco Pieri for their direction of my research.
\end{acknowledgments}

\bigskip 

\begin{thebibliography}{9}   

\bibitem{1} CMS PAS HIG-11-010 ($H\rightarrow\gamma\gamma$ PAS) - dsweb.cern.ch/record/1369553/files/HIG-11-010-pas.pdf
\bibitem{2} CMS PAS HIG-11-011 (Higgs combination PAS - cdsweb.cern.ch/record/1370076/files/HIG-11-011-pas.pdf
\bibitem{3} Other public plots - https://twiki.cern.ch/twiki/bin/view/CMSPublic/Hig11010TWiki

\end{thebibliography}

\end{document}